# Self-Consistent Ion Cyclotron Anisotropy-Beta Relation for Solar Wind Protons

Philip A. Isenberg[1], Bennett A. Maruca[2], and Justin C. Kasper[3]

[1]Institute for the Study of Earth, Oceans and Space, University of New Hampshire, Durham, NH 03824, USA; phil.isenberg@unh.edu
[2]Space Sciences Laboratory, University of California, Berkeley, CA 94720, USA; bmaruca@ssl.berkeley.edu
[3]Harvard-Smithsonian Center for Astrophysics, Cambridge, MA 02138, USA; jkasper@cfa.harvard.edu

**Abstract.** We derive a set of self-consistent marginally stable states for a system of ion-cyclotron waves propagating parallel to the large-scale magnetic field through a homogeneous proton-electron plasma. The proton distributions and the wave dispersions are related through the condition that no further ion-cyclotron resonant particle scattering or wave growth/damping may take place. The thermal anisotropy of the protons in these states therefore defines the threshold value for triggering the proton-cyclotron anisotropy instability. A number of recent papers have noted that the anisotropy of solar wind protons at 1 AU does not seem to be limited by the proton-cyclotron anisotropy threshold, even at low plasma beta. However, this puzzle seems to be due solely to the estimation of this anisotropy threshold under the assumption that the protons have a bi-Maxwellian distribution. We note that bi-Maxwellian distributions are never marginally stable to the resonant cyclotron interaction, so these estimates do not represent physically valid thresholds. The threshold anisotropies obtained from our marginally stable states are much larger, as a function of proton parallel beta, than the bi-Maxwellian estimates, and we show that the measured data remains below these more rigorous thresholds. Thus, the results of this paper resolve the apparent contradiction presented by the solar wind anisotropy observations at 1 AU: The bi-Maxwellian anisotropies are not rigorous thresholds, and so do not limit the proton distributions in the solar wind.

## 1. Introduction

The solar wind, among most other space plasmas, contains a significant population of ion-cyclotron waves (ICWs). These waves, which are the dispersive extension of the shear

Alfvén wave, have been observed in the solar wind for many years (Behannon 1976; Tsurutani et al. 1994; Jian et al. 2009; Jian et al. 2010; Podesta & Gary 2011; He et al. 2011; He et al. 2012), and presumably make up some fraction of the "slab component" of the turbulence in the wind (Matthaeus & Goldstein 1982; Bieber et al. 1996; Leamon et al. 1998; Dasso et al. 2005; Osman & Horbury 2009; Narita et al. 2010). Observations of solar wind ion distributions have been interpreted as resulting from resonant interactions between the ions and broad spectra of quasi-parallel ICWs, leading to ion scattering and heating (Marsch & Tu 2001; Marsch et al. 2004; Tu & Marsch 2002; Heuer & Marsch 2007; Matteini et al. 2007; Kasper et al. 2008, 2013; Bourouaine et al. 2010, 2011). Heating from ion-cyclotron dissipation may also be responsible for the observed perpendicular ion heating in the collisionless solar corona, which generates the fast solar wind (Hollweg & Isenberg 2002; Cranmer et al. 2007; Isenberg & Vasquez 2011), and for the preferential energization of heavy ions (Isenberg & Vasquez 2007, 2009).

This ion scattering and heating takes place through the resonant-cyclotron interaction, which can be described by the quasilinear formalism for small amplitudes and broad spectra of the waves (Kennel & Engelmann 1966; Montgomery & Tidman 1964; Stix 1992). The key relationship is the condition for cyclotron resonance

$$\omega(\mathbf{k}) - k_\parallel \, \upsilon_\parallel = \Omega , \qquad (1)$$

where the $\parallel$ ($\perp$) subscripts refer to components parallel (perpendicular) to the large-scale magnetic field $\mathbf{B}$. Condition (1) is satisfied when the frequency $\omega$ of a transverse wave with wavevector $\mathbf{k}$, Doppler-shifted into the reference frame moving along the magnetic field with particle speed $\upsilon_\parallel$, matches the gyrofrequency $\Omega = eB/mc$ of the particle. In this case, the particle and wave can exchange energy efficiently. A broad spectrum of randomly-phased waves interacting with a distribution of particles will lead to a stochastic diffusion of the resonant particles along characteristic paths in velocity space, as described in the next section. Such diffusion may yield a net heating or cooling of the particles, depending on the gradient of the phase-space particle density along these paths.

Of course, significant wave heating depends on the continual presence of ICWs in the face of dissipative losses. Since it is not currently known how or whether this wave power can be maintained in the corona or solar wind, the issue of ion heating by ICWs remains controversial.

However, the inverse process of anisotropy regulation by the ion-cyclotron anisotropy instability is expected to be highly effective, especially for low plasma $\beta$ (where $\beta$ is the ratio of particle thermal pressure to magnetic pressure). Ion distributions with large enough anisotropy (where we define the anisotropy as $A \equiv T_\perp / T_\parallel$) are unstable to the generation of ICWs, which propagate primarily along the magnetic field. This wave generation is accompanied by ion scattering to lower $v_\perp$, yielding a less anisotropic state. Thus, one expects the ions in a nearly homogeneous collisionless plasma, such as the *in situ* solar wind, to generally exhibit thermal anisotropies below the threshold value for exciting this instability.

In the past few years, a number of studies have been published which compile many years of solar wind data at 1 AU to investigate various statistical properties of the wind (Kasper et al. 2002, 2006, 2008, 2013; Hellinger et al. 2006; Bale et al. 2009; Maruca et al. 2011, 2012) (see also Marsch et al. (2004, 2006) and Matteini et al. (2007) for similar analyses at other radial positions). One output of these "metadata" studies which has received particular attention is the proton "anisotropy-beta" plot, an example of which is shown in Figure 1. This figure uses a subset of the solar wind core proton data from the SWE instrument on the WIND spacecraft collected near 1 AU between 1994 and 2010. The plot contains only data from times when the collisional age (defined by Kasper et al. (2008)) is $\tau_c \leq 0.1$, and the differential speed between the protons and alpha particles is < 0.8 times the Alfvén speed, $V_A$. When combined with standard criteria of data quality and spacecraft location, the observations plotted in Figure 1 comprise ~ 17% of the total solar wind data taken by SWE during this period. The contours in this plot show the occurrence probability in this nearly collisionless co-moving plasma of observing a given combination of proton thermal anisotropy and proton $\beta_\parallel$ in the core of the distribution. Similar versions of this figure, with slightly different data subsets and analysis procedures, can be found in earlier works (Hellinger et al. 2006; Bale et al. 2009; Maruca et al. 2011, 2012). The green and purple curves superimposed on the blue contours show the standard theoretical estimates of the threshold anisotropies for the proton-cyclotron anisotropy instability and the mirror instability, respectively. The green IC curve in particular is ultimately derived from the work of Gary et al. (1994), who investigated the growth/damping rates predicted for the parallel-propagating ion-cyclotron wave mode in a homogeneous bi-Maxwellian proton-electron plasma.

The data of Figure 1 seem to imply that the solar wind protons are not constrained by the proton-cyclotron anisotropy instability (Hellinger et al. 2006; Bale et al. 2009; Matteini et al. 2011; Maruca et al. 2011, 2012). This interpretation would be quite puzzling if it were true. However, as noted by Hellinger et al. (2006), the IC curve in the figure assumes bi-Maxwellian ion distributions, which may not be a good approximation in the solar wind. Recently, we pointed out that such an assumption is actually a very poor approximation when dealing with resonant wave-particle interactions in collisionless plasmas (Isenberg 2012, hereinafter Paper 1). In contrast to MHD fluid instabilities, such as the mirror or firehose which result primarily from anisotropies in the bulk particle pressures, resonant interactions depend strongly on the specific shape of the kinetic particle distribution. This shape changes significantly in response to the resonant scattering as the ions absorb or emit ICWs, and the self-consistent resonant interaction between collisionless ions and ICWs will never yield a bi-Maxwellian distribution. Equivalently, a bi-Maxwellian ion distribution can never be a state of marginal stability with respect to ICWs. Marginal stability signifies a state where the growth/damping rate of the waves (the imaginary part of the complex frequency when the wavenumber is taken real) $\mathrm{Im}(\omega) = 0$ for all values of $k$. The green curve in Figure 1 does not represent a rigorous threshold anisotropy, as given by the anisotropy of a marginally stable state. Instead, it is an estimate of how anisotropic an imposed bi-Maxwellian distribution must be to begin generating ICWs at some $k$ (while still damping them at other values of the wavenumber). The self-consistent result of such an instability will not be a bi-Maxwellian distribution of smaller anisotropy, but rather a stable distribution which cannot be accurately described by the bi-Maxwellian expressions.

In Paper 1, we derived a dispersion equation whose solution gave a self-consistent marginally stable proton distribution in the presence of parallel-propagating ICWs. This equation had a single free function, and we presented example solutions using a simple choice for this function. In this paper, we now set the free function to be Gaussian, a more physically realistic choice, and show that the threshold anisotropies obtained in this case are much higher than the bi-Maxwellian estimates previously quoted. These rigorous thresholds lie above all the solar wind data points in Figure 1, so the measured anisotropies do not in fact indicate any contradiction with the theoretical proton-cyclotron anisotropy instability.

In the next section, we briefly outline the concepts and formalism derived in Paper 1. In Section 3, we present results for the self-consistent marginally stable states based on Gaussian distributions, and put these results in the context of the anisotropy-beta measurements of the solar wind. Section 4 contains our further discussion and conclusions.

## 2. Self-consistent quasilinear ion-cyclotron interaction

Here, we sketch the mathematical formalism to be applied in this paper. A fuller, more detailed, exposition may be found in Paper 1.

In the quasilinear formalism, the resonant cyclotron wave-particle interaction can be determined from two conditions. The first is the condition for cyclotron resonance (1), specifying which wave modes resonate with which charged particles. The second condition describes the evolution of the particle distribution as a result of the interaction. Resonant particles diffuse along characteristic paths in velocity space determined from the dispersion relation of the waves. Since the particles must conserve their energy in the reference frame moving with the parallel phase speed of their resonant wave, these paths (actually surfaces) can be defined by

$$\eta^2 \equiv v_\perp{}^2 + v_\parallel{}^2 - 2\int_0^{v_\parallel} V_{ph}(v_\parallel{}')dv_\parallel{}' = \text{const.} \qquad (2)$$

where $V_{ph}(v_\parallel)$ designates that parallel resonant phase speed through equation (1). This new variable $\eta$, acts as a label for each resonant surface, and is determined by the value of $v_\perp$ for a given surface when $v_\parallel = 0$.

Resonant diffusion causes the particles to scatter down their density gradient along these surfaces. The quasilinear formalism conserves total energy in the system, so if this diffusion results in energy loss by the particles, the waves must gain the corresponding energy. Conversely, if the particle gradients are oriented such that the particles gain energy by diffusing along the resonant surfaces, this process is accompanied by wave damping. It follows that a particle distribution constructed with constant phase-space density over each resonant surface would be marginally stable with respect to the resonant cyclotron interaction: it would produce neither growth nor damping of the resonant waves. The dispersion relation describing the waves in this system would be purely real, containing no imaginary terms.

Such a construction is only possible if equation (2) describes a single set of nested surfaces, with no intersections or regions of overlap. The case we consider here, the resonant interaction between parallel-propagating ICWs and protons in a proton-electron plasma, has this property. More general situations can be more complicated. For instance, including heavy ions or off-axis wave propagation will yield surfaces which intersect over a finite region of $\upsilon_\parallel$, so such systems could not reach true marginal stability. However, the ion-cyclotron anisotropy instability produces predominantly parallel waves, and the question of whether this instability can regulate the anisotropy of solar wind protons can be addressed within this simpler framework. Systems with overlapping resonant surfaces can also be important for other aspects of wave-particle interactions (Isenberg & Vasquez 2007, 2009, 2011; Chandran et al. 2010).

The dispersion relation for parallel ICWs propagating through a homogeneous, collisionless plasma with otherwise arbitrary properties is well known (Montgomery & Tidman 1964) as

$$\omega^2 - c^2 k^2 + \pi\omega \sum_i \omega_{pi}{}^2 \int d\upsilon_\parallel \int \upsilon_\perp{}^2 d\upsilon_\perp \frac{\left(1 - \dfrac{k\upsilon_\parallel}{\omega}\right)\dfrac{\partial f_i}{\partial \upsilon_\perp} + \dfrac{k\upsilon_\perp}{\omega}\dfrac{\partial f_i}{\partial \upsilon_\parallel}}{\omega - k\upsilon_\parallel - \Omega_i} \;=\; 0 \qquad (3)$$

where the plasma frequency of the $i$th species of density $n_i$ is $\omega_{pi}{}^2 = 4\pi\, n_i\, q_i{}^2/m_i$ and $f_i$ is the normalized distribution function of that species. As detailed in Paper 1, we consider a proton-electron plasma in the limit of cold, massless electrons. We further apply charge neutrality and keep only the terms of lowest order in $V_A/c$. We also switch to normalized variables, taking the frequencies normalized to the proton cyclotron frequency $\Omega_p$, the wavenumbers normalized to the proton inertial length $V_A/\Omega_p$, and the speeds (including those in the definition of $\eta$) normalized to the Alfvén speed $V_A$. We then assume marginal stability, by taking the proton distribution function to be symmetric in $\upsilon_\parallel$ and constant along the resonant surfaces, so $f_p\,(\upsilon_\parallel,$ $\upsilon_\perp) = f\,(\eta)$. Under these circumstances, equation (3) reduces to the normalized equation for the self-consistent marginally stable state

$$\omega + k^2 - \pi k \int d\upsilon_\parallel \int \upsilon_\perp{}^3 d\upsilon_\perp \frac{\dfrac{\omega}{k} - V_{ph}(\upsilon_\parallel)}{\omega - k\upsilon_\parallel - 1} \frac{1}{\eta}\frac{\partial f}{\partial \eta} = 0 \;. \qquad (4)$$

At this point, neither $\omega(k)$ nor $V_{ph}(\upsilon_\parallel)$ are known and equation (4) must be solved self-consistently.

The dependence of the proton phase space density on the shell label, $\eta$, is the remaining free function referred to in the Introduction. In Paper 1, this function was taken as a simple box function, $f = S\,(\Gamma^2 - \eta^2)/N$, in order to facilitate the initial investigation of the solutions. There, $S(x)$ was the Heaviside step function, $N$ was the total proton density, and $\Gamma$ was a parameter giving the "thermal" width of the distribution.

We can consider this box function as the extreme case of a "thermal" proton distribution. We use this term here in a specific manner, to refer to a continuous function which has a single maximum at $\upsilon = 0$ and falls off with increasing proton energy. The box function defines the limit where this maximum is spread out over the entire flat distribution and only "falls off" at the zero of the step function. This distribution is clearly not realistic, and this paper presents results for a more physically reasonable shape of the proton distribution.

Here, we choose this free function to be a Gaussian:

$$f(\eta) = \frac{1}{N}\exp\left(-\frac{\eta^2}{\Gamma^2}\right) . \qquad (5)$$

With this choice, the integral over $\upsilon_\perp$ may be performed, resulting in

$$\omega + k^2 + \frac{2\pi\Gamma^2}{N}\int_0^\infty d\upsilon_\parallel \frac{\dfrac{\omega}{k} - V_{ph}(\upsilon_\parallel)}{\dfrac{\omega-1}{k} - \upsilon_\parallel}\exp\left(-\frac{\upsilon_\parallel{}^2 - 2\displaystyle\int_0^{\upsilon_\parallel} V_{ph}(\upsilon_\parallel{}')d\upsilon_\parallel{}'}{\Gamma^2}\right) = 0 . \qquad (6)$$

We obtain self-consistent numerical solutions of equation (6) for different values of $\Gamma^2$, through an iterative process as in Paper 1.

### 3. Results for Gaussian distribution

We obtain solutions of equation (6) for twelve values of $\Gamma^2$ and tabulate some properties of these marginally stable distributions in Table 1. With the normalized quantities defined above, the zeroth moment of $f$ gives the proton density $N$ in units of $V_A{}^3$, and the plasma betas are given by the respective second moments, $\beta_\parallel = <\upsilon_\parallel{}^2>/N$ and $\beta_\perp = <\upsilon_\perp{}^2>/(2N)$. We note that the combined relations (2) and (5) result in $\beta_\perp = \Gamma^2$.

The dispersion relations, $\omega(k)$, for four of the cases in Table 1 ($\Gamma^2 = 0.1$, 0.6, 4. and 9.) are shown in Figure 2, along with the ICW dispersion in a cold plasma. The ICW frequencies in these marginally stable plasmas approach the proton cyclotron frequency at smaller $k$ than in a cold plasma, and at somewhat smaller $k$ than the equivalent cases in Paper 1 as well. Thus, these waves have faster phase speeds than those predicted in the cold plasma case. Useful comparisons with bi-Maxwellian predictions are not possible since a bi-Maxwellian plasma does not allow any physically meaningful ICWs above a limiting value of $kV_A/\Omega \sim \beta_\parallel^{-1/6}$ (Stix 1992). The high-$k$ damping by bi-Maxwellian plasmas is so severe that numerical solutions tend to show Re($\omega$) decreasing with $k$ and eventually passing through zero (see e.g. (Yoon et al. 2010) for some examples). In contrast, the self-consistent marginally stable states in this paper support ICWs without growth or damping. The imaginary part of the frequency, Im($\omega$), is identically zero and undamped solutions extend to infinite $k$.

This identification of physically valid ICWs at large $k$ is a new regime for a thermal plasma. The limiting behavior in this regime can be obtained from equation (6) by taking $\omega \rightarrow 1$ as $k \rightarrow \pm\infty$ and considering the corresponding $v_\parallel$ and $V_{ph}(v_\parallel)$ to be small. Matching the second and third terms of equation (6) then yields

$$k^2 \approx \frac{2\pi\Gamma^2}{N} \frac{\omega}{|k|} \int_0^\infty \frac{dv_\parallel}{v_\parallel - \dfrac{\omega-1}{|k|}} \qquad (7)$$

leading to

$$\omega(k) \approx 1 - |k| \exp\left(-\alpha|k|^3\right) \quad (k \rightarrow \pm\infty) \qquad (8)$$

where $\alpha$ is proportional to $N/(2\pi\Gamma^2) = N/(2\pi\beta_\perp)$, a quantity available from the bulk fluid properties of the plasma.[1] We numerically verify the relation (8) for the different cases treated here and list the values of $\alpha$ for each case in Table 1. In Figure 3, we show the linear dependence of $\alpha$ on $N/(2\pi\Gamma^2)$, and find that

$$\alpha = 0.6846 \frac{N}{2\pi\beta_\perp}. \qquad (9)$$

---

[1] We note here that equation (17) in Paper 1, which is equivalent to (7) in this paper, has a typographical error. The lower limit on the integral in Paper 1 should be 0, not $-T$ as printed there.

This expression can be used to model the resonant cyclotron interaction for protons with small $v_\parallel$, within the thermal core of a distribution. In this limit, the resonant surfaces which direct the wave-particle scattering are given by equation (2) with

$$V_{ph}(v_\parallel) \approx -\text{sign}(v_\parallel) \left[ -\frac{\alpha}{\ln\left(|v_\parallel|\right)} \right]^{1/3} \quad (|v_\parallel| << 1), \qquad (10)$$

where $\alpha$ is given by equation (9).

The full self-consistent solutions determine the resonant surfaces throughout velocity space. In Figure 4, we show sample surfaces for the same cases shown in Figure 2. A specific surface is defined by the value of $\eta$ designating the place where that surface hits the $v_\parallel = 0$ axis, along with the value of $\beta_\parallel$. We show four sets of surfaces, corresponding to $\eta = 0.1, 0.2, 0.5,$ and 1. The surfaces are more anisotropic than the equivalent results from the cold plasma dispersion relation (also shown), and their anisotropy increases as $\eta$ decreases. Proton distributions with constant density along these surfaces will have higher anisotropies if they are concentrated near the origin, that is if they are colder. If a collisionless proton distribution has reached marginal stability with respect to the resonant cyclotron interaction with parallel-propagating ICWs, its density contours should follow the shape of these surfaces. It will be interesting to compare these shapes with the observed "quasilinear plateaus" reported by Marsch and co-workers (Marsch & Tu 2001; Tu & Marsch 2002; Heuer & Marsch 2007).

The anisotropies of these marginally stable proton distributions are identical, by definition, to the threshold anisotropies for the proton-cyclotron anisotropy instability. For the Gaussian dependence (5), these threshold anisotropies are listed in the last column of Table 1 for the cases computed here, and are plotted as a function of $\beta_\parallel$ in Figure 5. We find that these threshold anisotropies are well fit by the function

$$A = 2.16 + 1.56\,\beta_\parallel^{-0.343} \qquad (11)$$

which is shown by the black curve in Figure 5. We are not implying, by equation (11), that the theoretical threshold anisotropy should actually tend to 2.16 for very large $\beta_\parallel$, but only that this function fits the results in the range of $\beta_\parallel$ we have investigated. The points in Figure 5 are not well fit by the more commonly used function, $A = 1 + b\beta_\parallel^{-c}$. For comparison, the blue dashed curve in Figure 5 shows the function, $A = 1 + 0.43\beta_\parallel^{-0.42}$ derived by Gary et al. (1994) for bi-Maxwellian protons and quoted or rederived in many of the later observational works (Hellinger et al. 2006; Bale et al. 2009; Maruca et al. 2011, 2012). This curve is identical to the green "IC" curve in Figure 1. It is clear that, at least in the context of a cyclotron-resonant wave-particle interaction, the assumption of bi-Maxwellian particle distributions can yield results which are strongly misleading.

We also plot the function (11) as a red curve in Figure 1, labeled "msIC". It is seen that the marginally stable threshold anisotropies derived here lie well above the values measured in the solar wind. Thus, we find that there is no mystery regarding the action of the proton-cyclotron anisotropy instability in the solar wind. This instability only seemed to be inoperative due to the inaccurate estimates of the threshold obtained from the assumption of bi-Maxwellian proton distributions.

## 4. Discussion and Conclusions

In this paper, we derived a set of self-consistent plasma states containing an electron-proton plasma and parallel-propagating ICWs. These states are marginally stable with respect to the quasilinear resonant cyclotron interaction, with neither growth nor damping of the waves. In this sense, these states are time-stationary and may represent the time-asymptotic result of proton heating by a continuously driven spectrum of ICW power. We showed examples of the coupled wave dispersion and proton resonant surfaces for a range of proton plasma $\beta_\parallel$ corresponding to typical conditions in the solar wind.

We also determined the thermal anisotropies of these marginally stable states, which by definition give the threshold anisotropies for the proton-cyclotron anisotropy instability under out Gaussian assumption. We constructed a function (11), which fit the dependence of these anisotropies on the proton $\beta_\parallel$ and showed that these marginally stable thresholds lie well above the observed solar wind proton anisotropies at 1 AU.

We note that our analysis is still dependent on the free function, $f(\eta)$, which must be specified to solve equation (4). To the extent that these marginally stable states represent physical time-asymptotic states, this free function should have a "thermal" form, in the sense defined at the end of §2. Our choice of a Gaussian function in this paper seems to be the most plausible "thermal" choice to make, but other functional forms may also be important in the solar wind and other collisionless plasmas. At this point, we only have quantitative results for the box function case of Paper 1 and the Gaussian case given here, but some qualitative behavior of the solutions in general is evident.

First, we can see that the marginally stable dispersion curves, as shown in Figure 2, will be qualitatively similar independent of the choice of "thermal" $f(\eta)$. In particular, the functional form of $\omega$ as $k \to \pm\infty$ given by equation (8) follows from equation (4) for any "thermal" distribution which is smooth as $\eta \to 0$. Second, although the threshold anisotropy for the IC instability clearly depends on the choice of $f(\eta)$, the dependence is primarily due to the character of the resonant surfaces as illustrated in Figure 4. Ion cyclotron dispersion will always give proton resonant surfaces (as defined by equation (2)) which are nearly spherical for proton speeds $v >> V_A$, and become increasingly anisotropic for decreasing $v$. Thus, a choice of $f(\eta)$ which populates the inner (small $\eta$) surfaces more than the outer ones will yield a larger threshold anisotropy for a given $\beta_\parallel$. The box function of Paper 1 as an extreme case populated all the surfaces equally inside a given value, and it gave threshold anisotropies noticably larger than the bi-Maxwellian estimates of Gary et al. (1994). Our choice of a Gaussian in this paper is more physically reasonable, and it results in anisotropies well above the solar wind data. By this argument, proton distributions with high-energy tails, taking $f(\eta)$ as a kappa function for instance, would have somewhat smaller anisotropy thresholds, depending on the relative density in the high-energy tail. However, the exact thresholds obtained from such distributions have not yet been determined. It is likely that a fairly extreme high-energy tail would be needed to bring the anisotropy thresholds down to the level of the observed values.

Thus, we conclude that the puzzle of an ineffective proton-cyclotron anisotropy instability, represented by the green "IC" curve in Figure 1, can be completely resolved by recognizing that real collisionless protons will evolve away from a rigid bi-Maxwellian shape. In the solar wind at 1 AU, the proton anisotropy is still likely regulated by the mirror instability where $\beta_\parallel$ is high. At lower $\beta_\parallel$, the observed proton distributions are likely limited by the detailed effectiveness of the local perpendicular heating processes in the face of cooling by solar wind expansion. Since the observed anisotropies are not close to the marginally stable thresholds, it is not clear how much the measured proton distributions in the solar wind should actually be organized along the resonant surfaces shown in Figure 4. In fact, there may only be small differences between fits of measured proton distributions modeled as bi-Maxwellians compared to distributions modeled as partially filled resonant surfaces. Such comparisons will be the subject of future work.

Furthermore, space plasmas are often more complicated than a steady, homogeneous, proton-electron plasma, and additional plasma effects could also influence the threshold anisotropies. Matteini et al. (2011) suggested that an isotropic component of alpha particles would raise the bi-Maxwellian "IC" threshold to a more suitable level. Remya et al. (2013) have investigated the effects of anisotropic electrons in raising the threshold for ion-cyclotron instability in a number of space plasmas. Finally, Seough et al. (2013) point out that slow time variations of the magnetic field intensity in the solar wind can result in a statistical distortion or broadening of the effective bi-Maxwellian threshold. Such a broadening effect could appear as though the solar wind protons are not limited by a fixed threshold anisotropy. These suggested modifications may certainly affect the actual anisotropy thresholds in the solar wind, but the analysis behind all these suggestions currently remains based on the assumption of bi-Maxwellian particle distributions. We have seen here that such an assumption can be a very poor representation of collisionless space plasmas in general and the solar wind in particular.

This work was supported in part by NSF grants ATM0719738 and AGS 0962506, NASA grants NNX08AW07G and NNX11AJ37G, and DoE grant DEFG0207ER46372.

TABLE I.  Marginally stable proton distributions with Gaussian structure.

| $\Gamma^2 = \beta_\perp$ | $N$ | $\beta_\parallel$ | $\alpha$ | Anisotropy |
|---|---|---|---|---|
| 0.01 | $1.132 \times 10^{-3}$ | $3.976 \times 10^{-4}$ | $1.229 \times 10^{-2}$ | 25.15 |
| 0.03 | $7.566 \times 10^{-3}$ | $1.923 \times 10^{-3}$ | $2.709 \times 10^{-2}$ | 15.60 |
| 0.1 | $5.966 \times 10^{-2}$ | $1.032 \times 10^{-2}$ | $6.449 \times 10^{-2}$ | 9.694 |
| 0.36 | 0.5214 | $5.705 \times 10^{-2}$ | 0.1570 | 6.311 |
| 0.6 | 1.225 | 0.1099 | 0.2212 | 5.461 |
| 1.0 | 2.855 | 0.2082 | 0.3109 | 4.802 |
| 2.0 | 8.897 | 0.4829 | 0.4844 | 4.141 |
| 4.0 | 27.32 | 1.088 | 0.7439 | 3.677 |
| 5.0 | 39.09 | 1.405 | 0.8502 | 3.558 |
| 6.0 | 52.31 | 1.728 | 0.9490 | 3.472 |
| 7.0 | 66.88 | 2.056 | 1.041 | 3.404 |
| 9.0 | 99.69 | 2.723 | 1.207 | 3.306 |

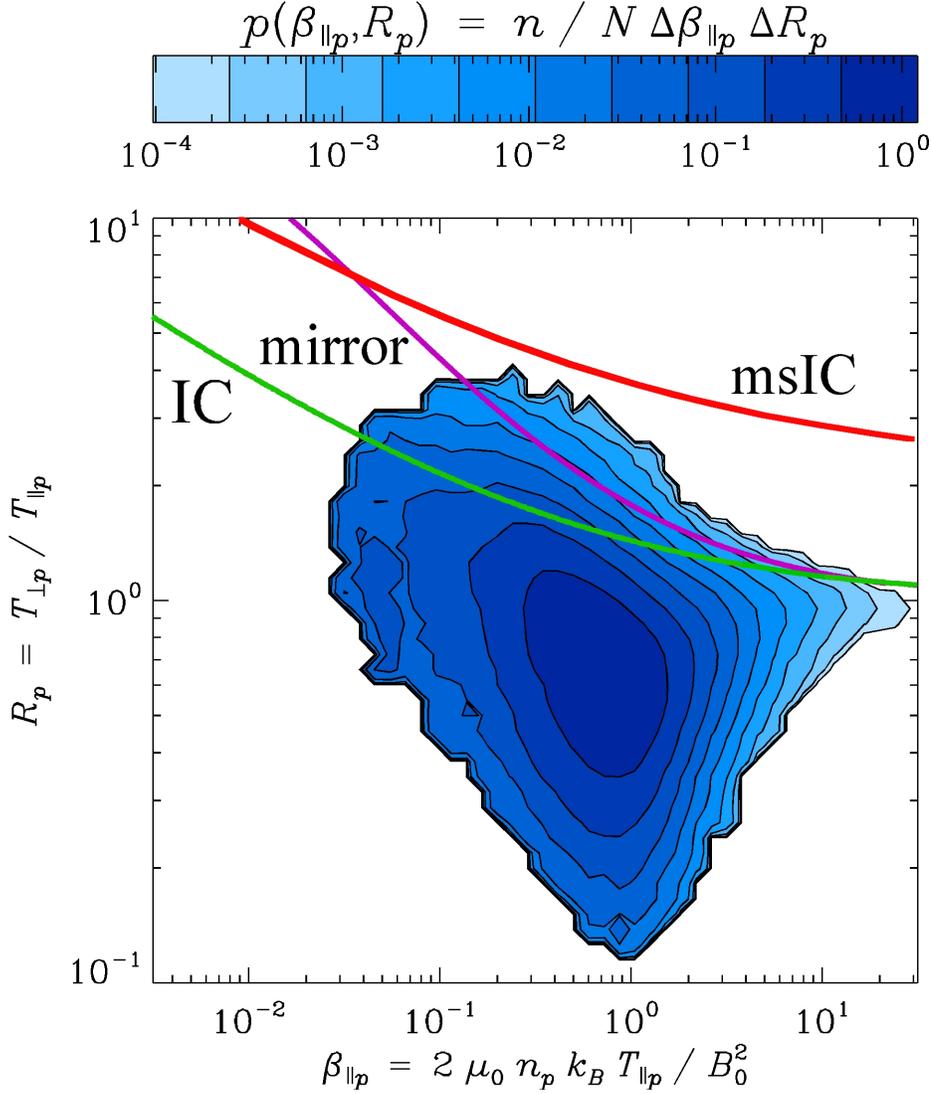

**Figure 1.** Relative probability of finding a given proton anisotropy and given proton $\beta_\parallel$ in 16 years of WIND/SWE data at 1 AU. These data are restricted to collisional age $\leq 0.1$ and differential speed between protons and alpha particles $< 0.8 \, V_A$. The green and purple curves overplotted on the blue contours indicate the traditional thresholds for triggering the ion-cyclotron anisotropy instability ("IC", green) and the mirror instability ("mirror", purple), estimated from analysis of bi-Maxwellian particle distributions. The red curve ("msIC") is from this work, and shows the ion-cyclotron anisotropy instability threshold determined from self-consistent marginally stable states of ICWs in a proton-electron plasma.

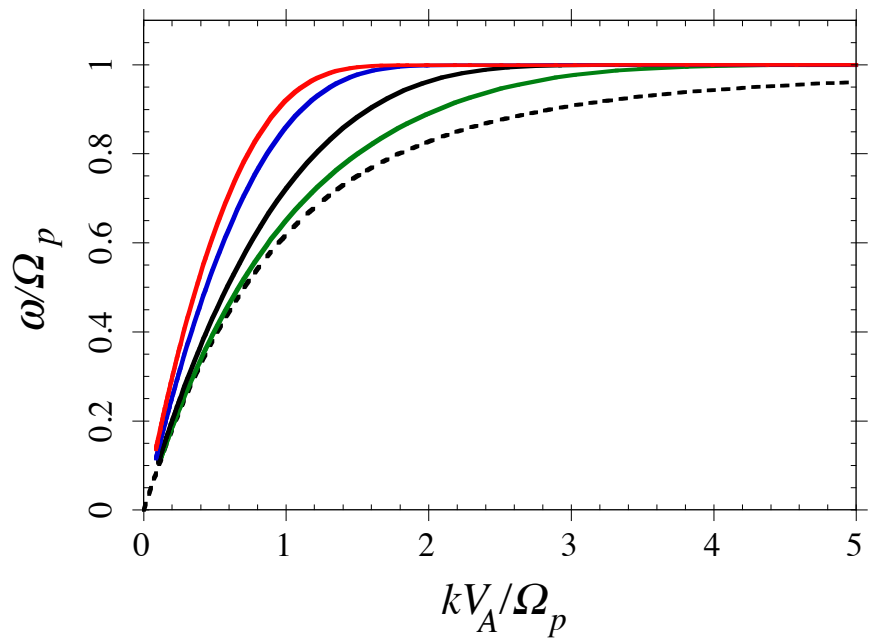

**Figure 2.** Dispersion curves for parallel-propagating ICWs in marginally stable plasmas for four values of $\beta_\parallel$: $\beta_\parallel = 0.01032$ (green), $\beta_\parallel = 0.1099$ (black), $\beta_\parallel = 1.088$ (blue), and $\beta_\parallel = 2.723$ (red). The black dashed curve shows the cold-plasma dispersion curve.

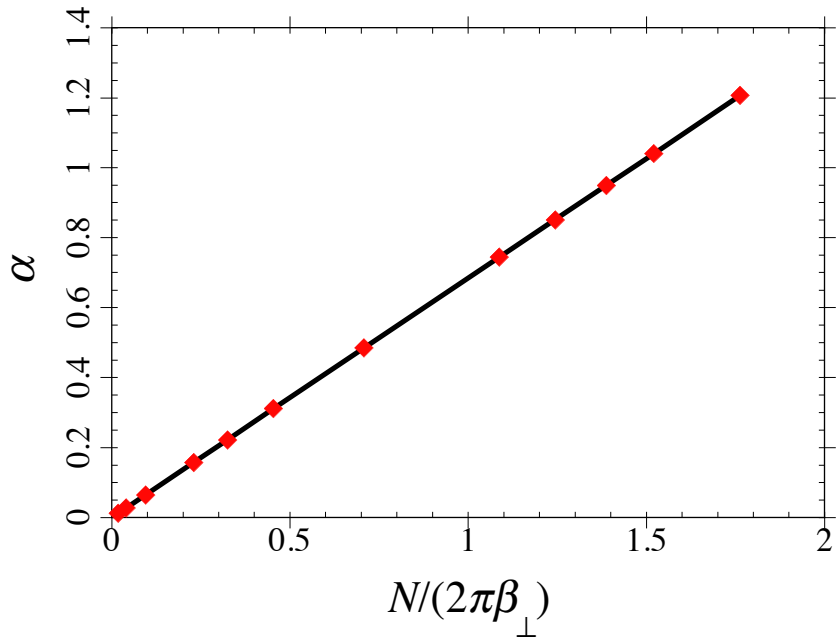

**Figure 3.** Plot of the parameter $\alpha$, which enters the expressions (8) and (10) for the dispersion and resonant surfaces at asymptotically high-$k$ values. The red diamonds show the numerically determined values given in Table 1. The black line shows the fit given in equation (9).

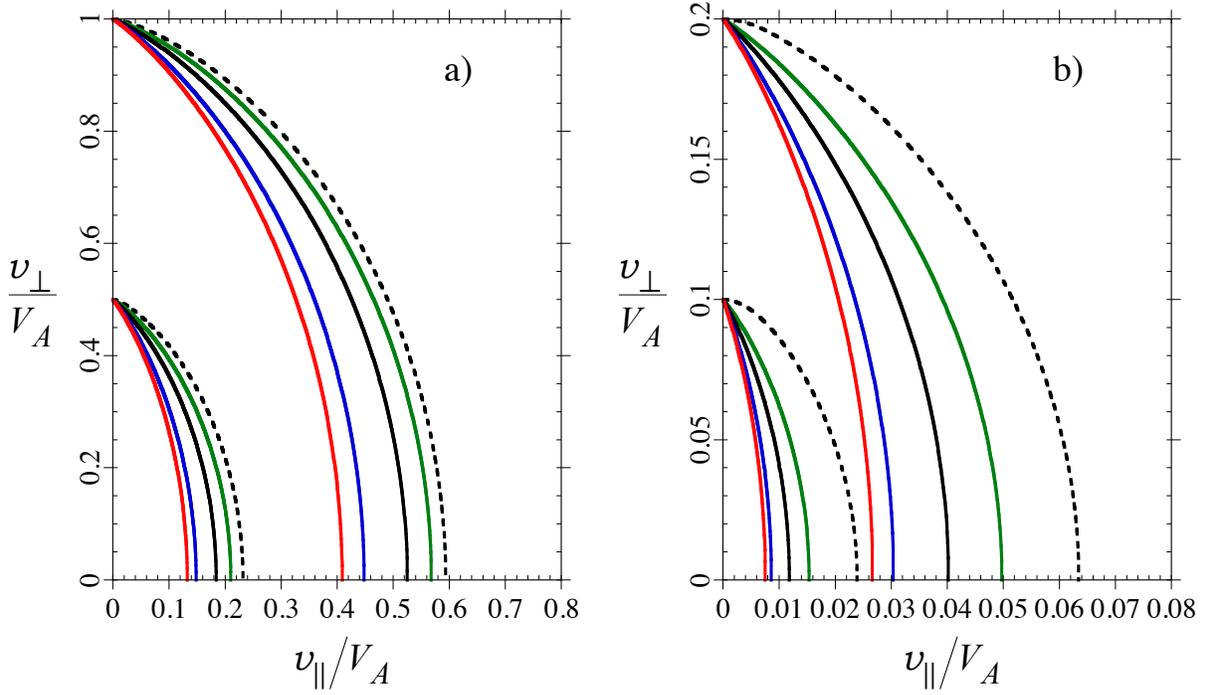

**Figure 4.** Representative resonant surfaces for protons in marginally stable states. For each value of $\beta_{\parallel}$ we show surfaces for four values of $\eta$, equivalent to their value of $\upsilon_{\perp}$ at $\upsilon_{\parallel} = 0$. The values of $\beta_{\parallel}$ and their color scheme are the same as in Figure 2: $\beta_{\parallel} = 0.01032$ (green), $\beta_{\parallel} = 0.1099$ (black), $\beta_{\parallel} = 1.088$ (blue), and $\beta_{\parallel} = 2.723$ (red). The black dashed lines show the equivalent surfaces using the cold plasma dispersion relation. a) $\eta = 0.5$ and 1.0. b) $\eta = 0.1$ and 0.2.

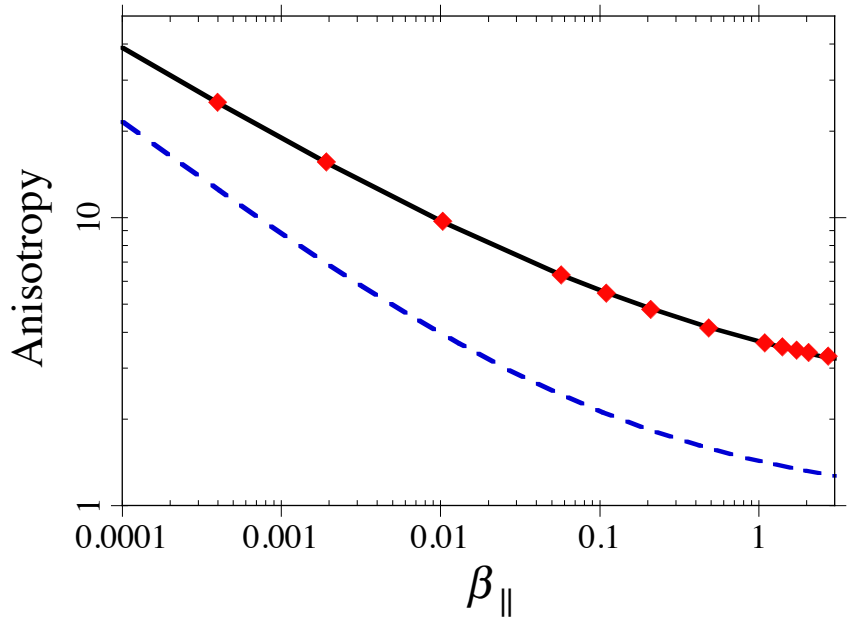

**Figure 5.** Marginally stable proton anisotropy, $T_\perp/T_\parallel$, as a function of the proton $\beta_\parallel$. The red diamonds show the numerically determined values given in Table 1, and the black curve shows the fit given in equation (11). The blue dashed curve shows the threshold estimated for a bi-Maxwellian proton distribution from Gary et al. (1994) if the maximum growth rate is taken at $10^{-3} \, \Omega_p$. This curve is identical to the green "IC" curve in Figure 1.